\documentclass[10pt, conference]{IEEEtran}
\IEEEoverridecommandlockouts
\usepackage{algpseudocode}                                 
\usepackage{algorithm}
\usepackage{graphicx}                                      
\usepackage{amsmath}
\usepackage{amssymb}
\usepackage{amsfonts}
\usepackage{amsthm}
\usepackage[mathcal]{eucal}
\usepackage{booktabs}
\usepackage{multirow}
\usepackage{color}
\usepackage[subrefformat=parens,farskip=0pt,justification=centering]{subfig}
\usepackage{color}
\usepackage{graphicx}
\usepackage{cite}                                          
\usepackage{comment}                                       
\usepackage{soul}                                          
\soulregister\cite7
\soulregister\ref7
\soulregister\pageref7
\usepackage{etoolbox}                                      
\usepackage{url}
\usepackage{nth}                                           
\usepackage{bm}                                            
\usepackage{courier}
\usepackage{balance}
\usepackage{threeparttable}
\usepackage[bookmarks=false]{hyperref}
\hypersetup{
    colorlinks = true,
    citecolor  = blue,
    linkcolor  = blue,
    urlcolor   = blue,
}
\usepackage{tikz}
\usetikzlibrary{positioning, fit}
\usepackage{filecontents}                                  
\usepackage{pgfplots}
\usepackage{pgfplotstable}
\pgfplotsset{compat=newest}
\usepackage{caption}
\usepackage{cleveref}
\Crefformat{figure}{Fig.~#2#1#3}                           
\Crefname{subfigure}{Fig.}{Figs.}
\Crefformat{table}{TABLE~#2#1#3}                           
\definecolor{CUHKorange}{RGB}{244,106,18} 
\definecolor{CUHKblue}{RGB}{0,111,190}    
\definecolor{CUHKgreen}{RGB}{0,127,128}   
\definecolor{CUHKred}{RGB}{228,46,36}     
\definecolor{CUHKyellow}{RGB}{198,148,34} 
\definecolor{CUHKdark}{RGB}{114,44,114}   
\definecolor{CUHKmiddle}{RGB}{144,44,144} 
\definecolor{CUHKlight}{RGB}{167,44,167} 
\definecolor{mygreen2}{RGB}{76,153,0}
\definecolor{mygrey}{RGB}{192,192,192}
\definecolor{myblue2}{RGB}{51,153,255}
\definecolor{myyellow}{RGB}{198,88,34} 
\definecolor{mygreen}{RGB}{134, 182, 54}
\definecolor{myorange}{RGB}{244, 106, 18}
\definecolor{mygray}{RGB}{102, 102, 102}
\definecolor{mymiddle}{RGB}{167, 44, 167}
\definecolor{mylight}{RGB}{191, 191, 255}
\definecolor{mydark}{RGB}{114, 44, 114}

\renewcommand{\vec}[1]{\boldsymbol{#1}}    

\algrenewcommand\textproc{\texttt}

\makeatletter
\let\OldStatex\Statex
\renewcommand{\Statex}[1][3]{%
  \setlength\@tempdima{\algorithmicindent}%
  \OldStatex\hskip\dimexpr#1\@tempdima\relax
}
\makeatother

\RequirePackage[normalem]{ulem} 
\RequirePackage{color}\definecolor{RED}{rgb}{1,0,0}\definecolor{BLUE}{rgb}{0,0,1} 

\graphicspath{{./figs/}}

\begin{document}

\title{
    CAD Tool Design Space Exploration via  \\ Bayesian Optimization
}

\author{\IEEEauthorblockN{Yuzhe Ma}
    \IEEEauthorblockA{
        Chinese University of Hong Kong\\
        \texttt{yzma@cse.cuhk.edu.hk}}
    \and
    \IEEEauthorblockN{Ziyang Yu}
    \IEEEauthorblockA{
        University of Hong Kong\\
        \texttt{yzy1996@connect.hku.hk}}
    \and
    \IEEEauthorblockN{Bei Yu}
    \IEEEauthorblockA{
        Chinese University of Hong Kong\\
        \texttt{byu@cse.cuhk.edu.hk}}
}

\maketitle

\begin{abstract}

The design complexity is increasing as the technology node keeps scaling down.
As a result, the electronic design automation (EDA) tools also become more and more complex. 
There are lots of parameters involved in EDA tools, which results in a huge design space. 
What's worse, the runtime cost of the EDA flow also goes up as the complexity increases, thus exhaustive exploration is prohibitive for modern designs.  
Therefore, an efficient design space exploration methodology is of great importance in advanced designs.
In this paper we target at an automatic flow for reducing manual tuning efforts to achieve high quality circuits synthesis outcomes.
It is based on Bayesian optimization which is a promising technique for optimizing black-box functions that are expensive to evaluate. 
Gaussian process regression is leveraged as the surrogate model in Bayesian optimization framework. 
In this work, we use 64-bit prefix adder design as a case study. 
We demonstrate that the Bayesian optimization is efficient and effective for performing design space exploration on EDA tool parameters,
which has great potential for accelerating the design flow in advanced technology nodes.

\end{abstract}

\section{Introduction}
Electronic design automation (EDA) tools play a vital role in pushing forward the VLSI industry.
Nowadays, the design complexity keeps increasing as the technology node scales down.
As a result, EDA tools correspondingly become more and more complex since more sophisticated algorithms and optimizations are incorporated to ensure timing closure, reliability and manufacturability.
Essentially, the increasing complexity corresponds to the expanding amount of parameters involved in EDA tools, which implies a huge design space and requires rich expertise from the designers to achieve desired quality. 
Due to the underlying complicated optimization process, it is commonly seen that a subtle change of constraints would lead to large variations in final design performances, which makes designers could not rely on the intuition to explore the design space. 
What's worse, the runtime cost of synthesis flow also goes up as the complexity increases, and the exhaustive exploration is prohibitive for modern designs.  
Therefore, an efficient design space exploration methodology is of great importance in advanced designs.

Take the synthesis and physical design flow as an example. 
Given a specific design and a set of CAD tool scripts, parameterized by a vector representation $\vec{x}$, the synthesis flow is performed and the core metrics, including area, power, and delay, can be obtained. 
The aim is to find the most suitable parameters in CAD tool scripts that result in the best circuit performance after synthesis. 
Intuitively, it can be formulated as an optimization problem. 
However, the synthesis process is too sophisticated to be analytically modeled. 
Also, the solution space is too large.
Therefore, designers can neither derive a closed-form solution, nor exhaustively search the solution space.
Instead, the optimization should be conducted in an exploration manner in the design space. 

Data-driven approaches like machine learning and deep learning have been heavily applied in EDA field,
including testability analysis, physical design, mask optimization and so on \cite{TEST-DAC2019-Ma,FPGA-ICCAD2017-Pui,OPC-TCAD2020-Yang}. 
Besides these typical stages in design flow, there is a rich literature investigating how to perform design space exploration effectively and efficiently with learning-based methods. 
Linear regression and artificial neural network are adopted for multiprocessor systems-on-chip design \cite{CAD-TCAD2009-Palermo}. 
Random forest is applied in high level synthesis (HLS) \cite{CAD-DAC2013-Liu,DSE-DATE2016-Meng}. 
Roy \textit{et al}.~propose to explore power efficient adders by leveraging support vector machine (SVM) to predict the power-performance-area (PPA) values \cite{DSE-ISLPED2017-Roy}. 
Beyond that, a more efficient design space exploration flow for high performance adders is introduced in \cite{DSE-TCAD2018-Ma}, which is based on an active learning flow with Gaussian process regression as the surrogate model.

Bayesian optimization is a promising technique for optimizing black-box functions that are expensive to evaluate, which promises significant automation such that both design quality and human productivity can be greatly improved.
It has been applied in designing various sorts of circuits. 
A DNN accelerator is designed using Bayesian optimization for design parameters search \cite{DSE-ISLPED2017-Reagen}. 
Recently, Bayesian optimization has been heavily applied in analog circuits design as well \cite{DSE-ICML2018-Lyu,DSE-DATE2019-Zhang,DSE-DAC2019-Zhang}, in which several approaches are introduced to improve the efficiency.  
Specifically, Lyu \textit{et al}.~propose to compose a multi-objective acquisition function to enhance the efficiency of the sampling step \cite{DSE-ICML2018-Lyu}. 
In \cite{DSE-DATE2019-Zhang}, neural networks are leveraged to generate the feature representation explicitly, which reduce the prediction time complexity from $\mathcal{O}(n^2)$ of traditional kernel function-based computation to $\mathcal{O}(1)$. 

\begin{figure*}[!tb]
	\begin{minipage}{0.65\linewidth}
		\centering
        \hspace{-.2in}
        \subfloat[]{ \includegraphics[width=0.66\textwidth]{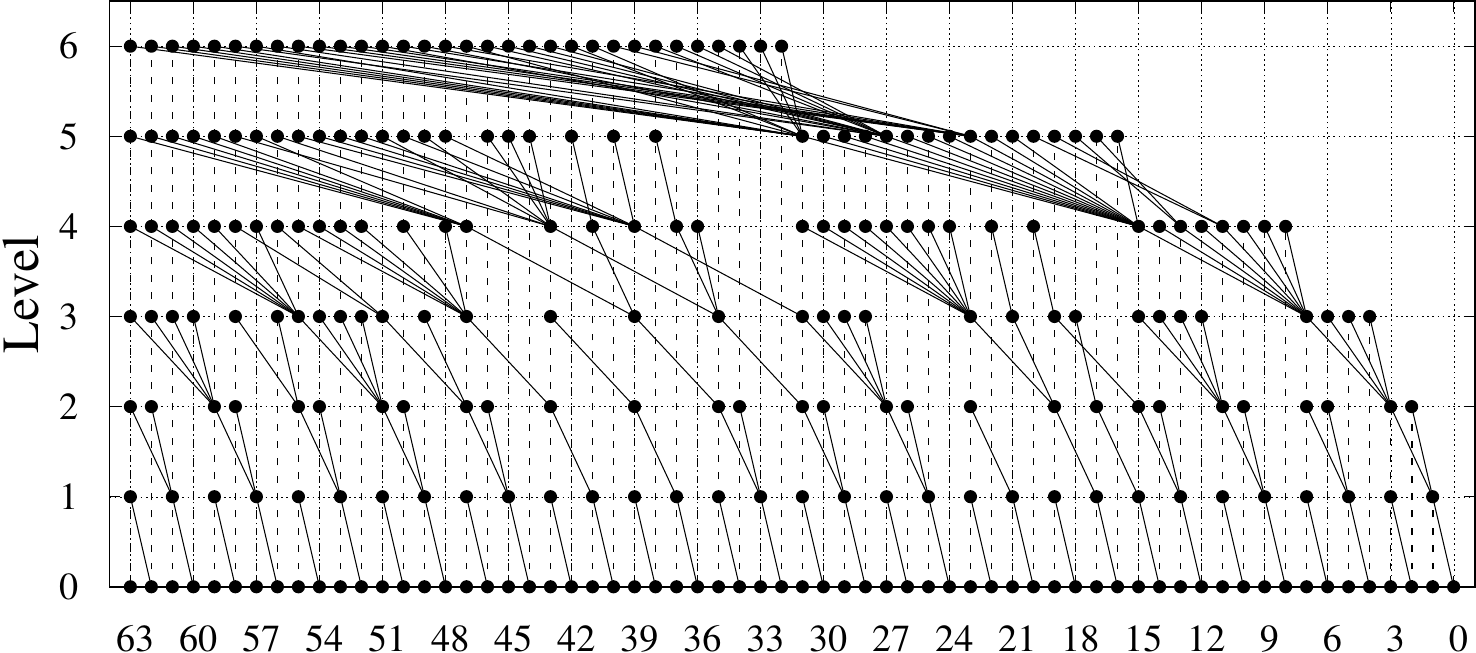} 
        \label{fig:arch-front}} \hspace{.01in}
        \subfloat[]{ \includegraphics[width=0.32\textwidth]{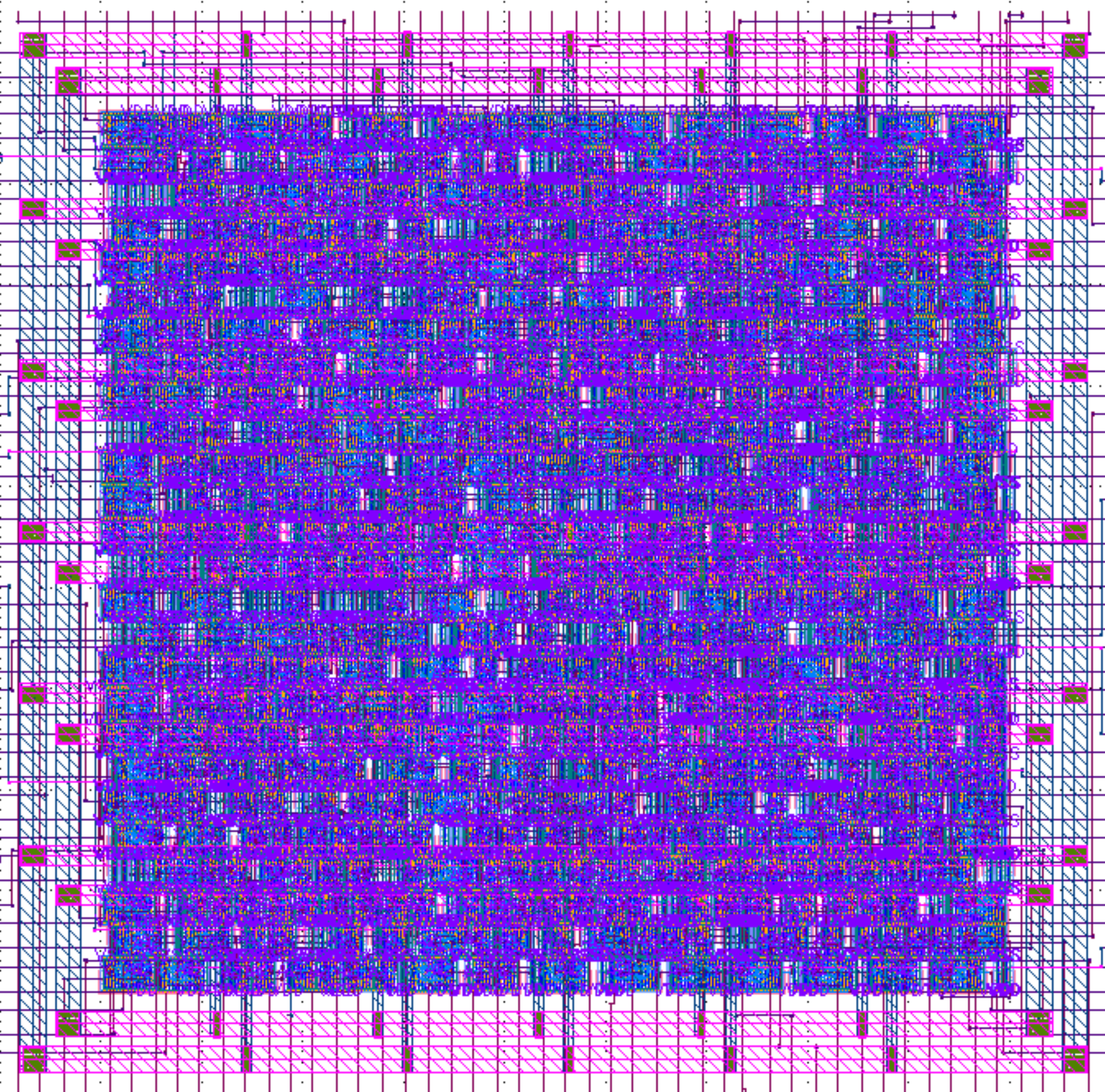} \label{fig:arch-back}}
        \caption{(a) An example of architectural solution; (b) Corresponding physical solution.}
        \label{fig:arch}
	\end{minipage}
    \hspace{.2in}
	\begin{minipage}{0.35\linewidth}
        \centering
        \includegraphics[width=.88\textwidth]{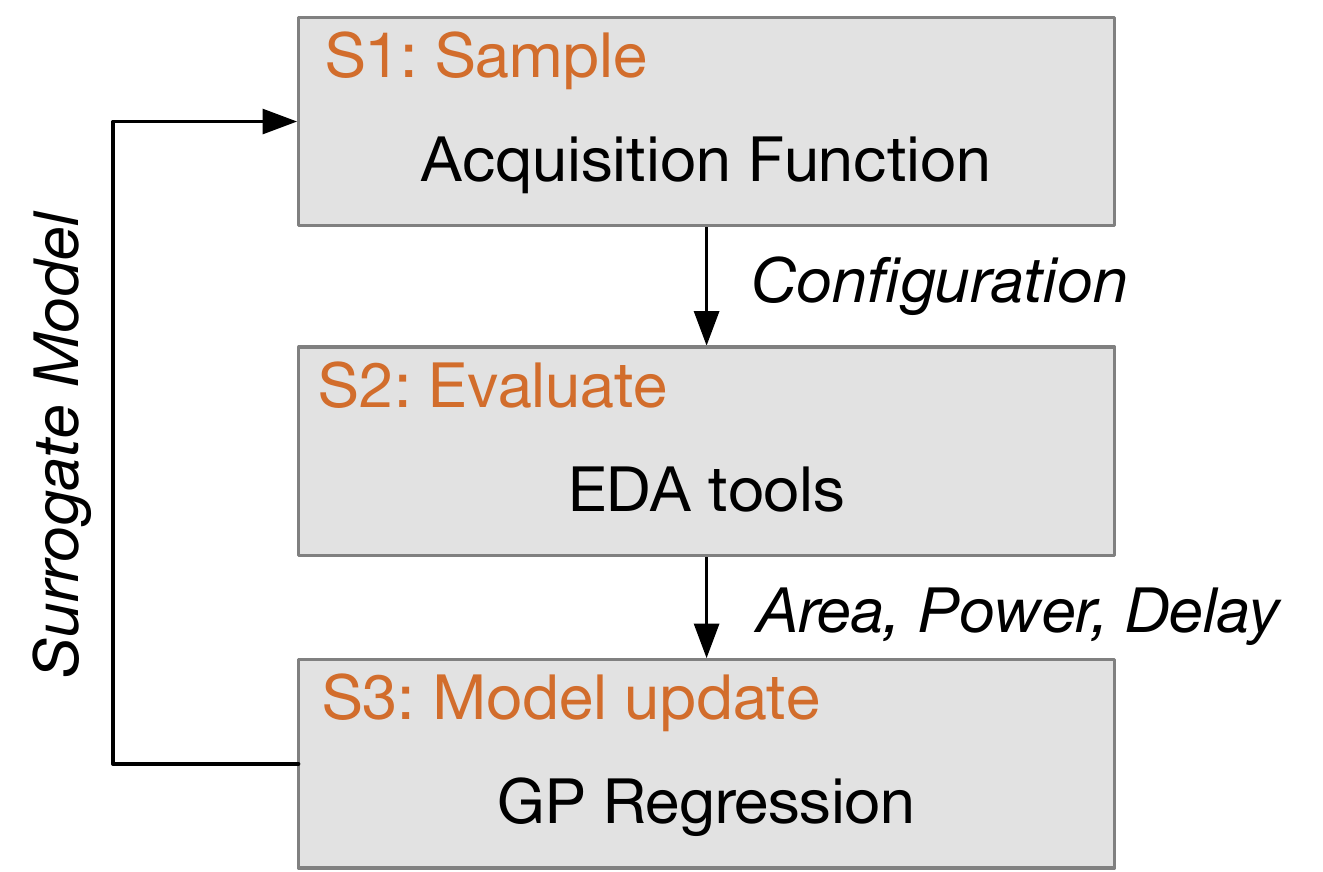}
        \caption{Diagram of the proposed workflow.}
        \label{fig:flow}
	\end{minipage}
\end{figure*}

In this paper we target at an automatic flow for reducing manual tuning efforts to achieve high quality circuits synthesis outcomes. 
We use 64-bit adder design as a case study. 
\Cref{fig:arch} demonstrates two significantly different views of the same implementation of a regular 64-bit Sklansky adder, in which \Cref{fig:arch-front} is the prefix structure in the front-end, and \Cref{fig:arch-back} is the result after this design goes through the back-end design process. 
Given a prefix architectural design, how to obtain efficiently a result of high quality after back-end design is a critical problem. 
There exist two main challenges. 
On one hand, it is observed that the final performance after logical synthesis and physical synthesis could vary significantly under subtle change of sensitive constraints. 
On the other hand, the design space is huge and it grows exponentially as more constraints are taken into consideration, which makes an efficient flow in high demand. 
To tackle these issues, we develop a Bayesian optimization-based approach for efficient exploration in the huge design space. 
Gaussian process regression (GPR) is leveraged as the surrogate model in our Bayesian optimization framework, which captures the correlation among the training samples.
Based on the surrogate model, different acquisition functions are investigated to guide the exploration process to find the superior points.
Various regular adder structures including legacy regular structures, synthesized structures and structures from commercial IP library are used for experimental evaluation. 
We demonstrate the Bayesian optimization is efficient and effective for performing design space exploration on CAD tool parameters, which has great potential for accelerating the design flow in advanced technology nodes. 

The rest of this paper is organized as follows. 
Preliminary knowledge on Bayesian optimization and tool settings are introduced in \Cref{sec:prelim}. 
\Cref{sec:algo} depicts the proposed design flow using Bayesian optimization. 
Experimental evaluations are presented in \Cref{sec:results}, and \Cref{sec:conclu} concludes the paper.

\section{Preliminaries}
\label{sec:prelim}
\subsection{Bayesian Optimization}

Essentially, Bayesian optimization is a statistical framework to optimize an arbitrary objective function. 
It is able to take advantage of the full information gained from past experiments to make the search much more efficient than human tuning. 
The core components of Bayesian optimization consist of a probabilistic surrogate model which is cheap to evaluate and provides our beliefs about the black-box function $f(\cdot)$, and an acquisition function which is designed to select the most informative one to query the label, such that the optimizer can be attained with a minimal number of expensive evaluations on the objective functions. 
After observing the output $f(\vec{x})$ of each new set of parameters $\vec{x}$ in each iteration, the surrogate model is updated such that it can better model the posterior distribution over the space of the objective function. 

\subsection{Gaussian Process Regression}
The surrogate model plays a vital role in the Bayesian optimization framework. 
Gaussian Process (GP) model is typically adopted in building the surrogate models thanks to its capability to capture the uncertainty of the prediction. 
A Gaussian process is specified by its mean function and covariance function. 
Conventionally, the training process selects the parameters in the presence of training data such that the marginal likelihood is maximized. 
Then the Gaussian Process model can be obtained and the regression can be proceeded with supervised input \cite{GP-B2006-Rasmussen}. 
A GP learner provides a Gaussian distribution $\mathcal{N} $ of the values predicted for any test input $\vec{x}$ by computing
\begin{equation}
\begin{aligned}
m(\vec{x})             & = k(\vec{x},\mathbf{X})^{\top}(k(\mathbf{X},\mathbf{X})+\sigma_n^{2}\mathbf{I})^{-1}\mathbf{Y},\\
\sigma^{2}({\vec{x}})  & = k(\vec{x},\vec{x}) -
k(\vec{x},\mathbf{X})^{\top}(k(\mathbf{X},\mathbf{X})+\sigma_n^{2}\mathbf{I})^{-1}k(\vec{x},\mathbf{X}),
\end{aligned}
\label{eq:gp}
\end{equation}
where $\mathbf{X}$ is the training set, $\mathbf{Y}$ is the supervised information of trained set $\mathbf{X}$. 
In the case of CAD tool design space exploration, a prediction of a design objective consists of a mean and a variance from Gaussian Process regression.
The mean value $m(\vec{x})$ represents the predicted value and the variance $\sigma(\vec{x})$ represents the uncertainty of the prediction. 
To make sure the model is close to the actual situation, additive noise $\varepsilon$ is introduced on the black-box function value $z$.
It is assumed that the noise is Gaussian distributed: $\varepsilon \sim \mathcal{N}(0, \sigma_n^2)$. 
The noisy observed value is: $y = z + \varepsilon$.
$k$ is covariance function characterizes the correlations between different samples in the process. 

\subsection{Tool Settings}

\begin{table}[!tb]
	\centering
	\caption{Design Parameters in CAD Tools}
	\label{tab:params}
    \resizebox{8.6cm}{!}{
    \renewcommand{\arraystretch}{1.1}{
	\begin{tabular}{c|c|c|c|c}
		\toprule
		Parameter & Min. & Max. & Type & Stage \\ \midrule
		\texttt{max\_delay}          & 0.1   & 0.5   & float   & LS \\
		\texttt{clock\_period}       & 1.0   & 2.0   & float   & LS \\
		\texttt{pin\_load}           & 0.002 & 0.006 & float   & LS \\
		\texttt{output\_delay}       & 0.1   & 0.5   & float   & LS \\
		\texttt{core\_utilization}   & 0.5   & 1.0   & float   & PD \\
		\texttt{core\_aspect\_ratio} & 1     & 3     & integer & PD \\
		\bottomrule
	\end{tabular}
    }
    }
\end{table}

The parameters in design space is depicted in \Cref{tab:params}, including both the logical synthesis (LS) and the physical design (PD) stages. 
The data types are not fixed and can be either floating value or integer value, depending on the specific parameters. 
These parameters essentially set constraints on the area, power and timing during the synthesis process, thus have great impacts on the ultimate performances. 

\subsection{Problem Definition}
The design space exploration for CAD tools can be treated as an optimization problem.
The objective is to minimizing the performance-power-area (PPA) values of a design, subject to that a set design constraints is satisfied.
The variables of the problem are also constrained by a bounded set which indicates reasonable ranges of the tool parameters.

\section{Methodology and Design Flow with \\ Bayesian Optimization}
\label{sec:algo}
\subsection{Design Flow}

\Cref{fig:flow} depicts the proposed iterative optimization flow. 
Each pass consists of a 3--step process and finally produces one sample point. 
In each iteration, it begins with solving the acquisition function, which returns a set of parameters. 
Since acquisition function is to guide the choice of the next most potential candidate sample, the choice of it depends on the specific case. 
A commonly used acquisition function is upper confidence bound (UCB) \cite{GP-ICML2010-Srinivas}, which is defined as	
\begin{equation}
    \operatorname{UCB}(\boldsymbol{x})=m(\boldsymbol{x})+\kappa \sigma(\boldsymbol{x}),
    \label{eq:ucb}
\end{equation}	
where $m(\boldsymbol{x})$ and $\sigma(\boldsymbol{x})$ are the predictive value and uncertainty of GP defined as \Cref{eq:gp}, respectively.
$\kappa$ is a parameter that balances the exploitation and exploration. 
UCB can be considered as the weighted sum of the predictive value and uncertainty. 
When the weight $\kappa$ is small, UCB tends to help select sample with low expected value. 
While with a large $\kappa$, this acquisition function is more likely to select samples with large uncertainty. 

In the case of CAD tool design space exploration, the final objective functions are to be minimized, which is in conflict with UCB. 
Instead, lower confidence bound (LCB) function is leveraged as the acquisition function, which is defined as follows:	
\begin{equation}
\operatorname{LCB}(\boldsymbol{x})=m(\boldsymbol{x})-\kappa \sigma(\boldsymbol{x}).
\label{eq:lcb}
\end{equation}	

Besides LCB, several other acquisition functions are also commonly used. 
Probability of improvement (POI) evaluates the objective function $f$ and finds the sample which is the most likely to attain improvements upon the best observed value $f^*$:
\begin{eqnarray}
    \begin{split}
        \operatorname{POI}(\boldsymbol{x})
        &= \operatorname{P}(f(\boldsymbol{x})\leq f^*-\zeta) \\
        &= \Phi(\frac{m(\boldsymbol{x})-f^*+\zeta}{\sigma(\boldsymbol{x})}).
    \end{split} 
\end{eqnarray} 
Here $\Phi$ is the normal cumulative distribution function (CDF). 
$\zeta$ is a small trade-off parameter. 
Searching with POI ignores the margin of improvement and only selects those samples with higher probability of improving even if the improvement is smaller, which is prone to stucking at local optima.

An alternative acquisition function is expected improvement (EI). 
EI incorporates the margin of improvement by maximizing the expectation of the improvement. 
The improvement compared to the best observed sample can be written as:
\begin{eqnarray}
\operatorname{I}(\boldsymbol{x}) = \max({0,f^*-f(\boldsymbol{x})-\zeta}).
\end{eqnarray}
Assuming the value of objective function $f(\boldsymbol{x})$ is normally distributed, $y \sim \mathcal{N}(m(\boldsymbol{x}), \sigma^2(\boldsymbol{x}))$, EI could be written as below:
\begin{eqnarray}
\begin{split}
\operatorname{EI}(\boldsymbol{x}) &= \operatorname{E}(\operatorname{I}(\boldsymbol{x}))\\
&= (f^*-m(\boldsymbol{x})-\zeta)\Phi(\frac{f^*-m(\boldsymbol{x})-\zeta}{\sigma(\vec{x})})\\
&+\sigma(\vec{x})\phi(\frac{f^*-m(\boldsymbol{x})-\zeta}{\sigma(\vec{x})}).
\end{split}
\end{eqnarray}
Here $\phi$ is the standard normal probability distribution function (PDF). 
$\zeta$ is the parameter set to trade off between exploration and exploitation.

After obtaining a set of parameters, the second step is to conduct the synthesis flow to obtain the corresponding performance of the target design under selected parameters. 
The synthesis flow involves logic synthesis and physical design, which makes this step the most time-consuming among all three steps. 
Typically, EDA tools are launched using $\mathsf{Tcl}$ scripts. 
Therefore, the selected parameters need to be translated into a representation so that the EDA tools can interpret.
To do so, a $\mathsf{Tcl}$ script generator is built to provide such scripts based on the parameters. 

The last step in each iteration is to update GP regression model using the evaluation result, i.e., area, power and delay values, which is GP model training. 
Assuming $n$ targeted values $\vec{y}$ are already evaluated from the training set $\mathbf{X}$. The joint distribution of the $\vec{y} \in \mathbb{R}^n$ and the test sample value $z$ can be written below:
\begin{eqnarray}
    \begin{bmatrix}
        \vec{y}\\
        z
    \end{bmatrix} 
    \sim \mathcal{N}\left (0,
    \begin{bmatrix}
        k(\mathbf{X},\mathbf{X})+\sigma_n^{2}\mathbf{I} & k(\mathbf{X}, \vec{x})\\
        k(\vec{x}, \mathbf{X}) & k(\vec{x},\vec{x})
    \end{bmatrix} \right).
\end{eqnarray}

The mean function $m(\vec{x})$ and covariance function $\sigma^2(\vec{x})$ in \Cref{eq:gp} can be derived from the above equation. 
	
We optimize the GP regression model by maximizing the marginal likelihood $p(\vec{y}|\mathbf{X},\vec{\mu})$. Here $\vec{\mu}$ represents the vector of all parameters contained in the model. This marginal likelihood is the marginalization over the black-box function values $\vec{z}$. From Bayesian theory, this can be represented below:
\begin{eqnarray}
p(\vec{y}|\mathbf{X},\vec{\mu}) = \int p(\vec{y}|\vec{z},\mathbf{X},\vec{\mu})p(\vec{z}|\mathbf{X},\vec{\mu})d\vec{z}.
\end{eqnarray} 
In GP regression model, the likelihood term $p(\vec{y}|z,\mathbf{X},\vec{\mu})$ and prior $p(z|\mathbf{X},\vec{\mu})$ are all Gaussian. Finally we could derive the log marginal likelihood:
\begin{eqnarray}
\begin{split}
\log p(\vec{y}|\mathbf{X},\vec{\mu}) = &-\frac{1}{2}\vec{y}^\top(k+\sigma_n^2\mathbf{I})^{-1}\vec{y} - \frac{1}{2}\log|k+\sigma_n^2\mathbf{I}|\\
&-\frac{n}{2}\log 2\pi .
\end{split}
\end{eqnarray}
The parameters in the model are updated after the log marginal likelihood are maximized in every iteration.	
Then the model can be leveraged by acquisition function to explore new point in the design space.

\subsection{Multi-objective Optimization with Bayesian Optimization }

The paradigm of conventional Bayesian optimization is natural for optimizing a single objective since the acquisition function is assumed to evaluate a single value. 
Regarding the design space exploration in circuits design, the most conventional objective is a vector of performance metrics, i.e., PPA. 
There exists strong trade-offs among these metrics in practice. 
Therefore, dedicated strategies are required to adopt Bayesian optimization to handle this problem. 
A straightforward way is to optimize a single metric value in the performance metric vector at a time in one Bayesian optimization procedure and select several better designs on each metric only. 
Then the whole Bayesian optimization procedure is repeated for multiple times, and each procedure finds superior designs on different metrics. 
The Pareto optimal ones in the PPA objective space can be picked manually by merging these selected designs. 

Alternatively, scalarization can be performed to transform the three dimension vector of this three-dimensional space (delay vs.~power vs.~area) into a joint output as the regression target rather than using any single output \cite{DSE-TCAD2018-Ma,LEARN-ICPR1996-Tumer}, which is formulated as  
\begin{equation}
	PPA = \alpha_1 \cdot Area + \alpha_2 \cdot Power + Delay.
\end{equation}
Scalarization of the three metrics provides a weighted and linear relation among them.
Intuitively, by changing the weight values, the Gaussian regression model will try to maximize the prediction accuracy on the most weighted direction. 
On the contrary, other metric directions will be predicted with less accuracy hence introducing relaxations to some extent.  
Therefore, we can actually explore a much larger space if we sweep the weight values over a wide range from 0 to large positive values \cite{DSE-ISLPED2017-Roy}.
The regression model will be guided to explore different best solutions which are Pareto-optimal. 
Once it is obtained, the designers can just select the most suitable designs according to specific constraints.

\section{Experimental Results}
\label{sec:results}
We perform comprehensive experiments to validate the proposed Bayesian optimization approach for automatic design space exploration in back-end design flow. 
The experiments are conducted on a 2.2GHz Linux machine with 32GB memory. 
We use Design Compiler \cite{TOOL-dc} (version F-2014.09-SP5) for logical synthesis, and IC Compiler \cite{TOOL-icc} (version N-2017.09) for the placement and routing.
``\texttt{tt1p05v125c}'' corner and Non Linear Delay Model (NLDM) in 32$nm$ SAED cell-library for LVT class \cite{LIB-SAED} (available by University Program) is used for technology mapping.
The core part in Bayesian optimization engine is implemented with $\mathsf{Python}$. 
Within the engine, $\mathsf{Tcl}$ scripts are generated automatically based on the explored parameters, which are used to launch the back-end design tools. 
We use various adder designs, including regular structures like Sklansky adder and Kogge-Stone adder, synthesized adder structure obtained from \cite{DSE-TCAD2018-Ma},
as well as the designs from commercial DesignWare IP libraries .

\subsection{Results Comparison with Industrial Settings}

\begin{table*}[!tb]
	\centering
	\caption{Performance comparison between BO and baseline settings on single objective}
	
	\label{tab:bl-bo-single}
    \resizebox{13cm}{!}{
    \renewcommand{\arraystretch}{1.2}{
	\begin{tabular}{c|cc|cc|cc}
		\toprule
		Adder structure     &\multicolumn{2}{c|}{ Area ($\mu m^2$)} &\multicolumn{2}{c|}{Energy ($fJ/op$)} &\multicolumn{2}{c}{Delay ($ps$)}\\ 
		\cline{2-7}       & Baseline & BO & Baseline & BO & Baseline & BO \\ 
		\hline
		\texttt{DesignWare}  & 2531 & 1873 & 8160 & 6220 & 346 & 328 \\
		\texttt{Sklansky}    & 1792 & 1772 & 6100 & 5020 & 356 & 350 \\
		\texttt{Kogge-Stone} & 2563 & 2323 & 8780 & 6630 & 347 & 338 \\
		\texttt{Synthesized} & 1753 & 1740 & 5900 & 5000 & 353 & 345 \\
		\hline
		Average           & 2160.1 & 1927.4 & 7235.0 & 5717.5 & 350.9 & 340.7 \\
		Ratio             & 1.0    & 0.89   & 1.0    & 0.79   & 1.0   & 0.97  \\
		\bottomrule
	\end{tabular}
    }
    }
\end{table*}

\begin{figure}[tb!]
    \centering
    \subfloat[]{\pgfplotsset{
    width=7.8cm,
    height=4.2cm
}

\begin{tikzpicture}
    \begin{axis}[
            ybar,
            xticklabels={\texttt{Sklansky}, \texttt{Synthesized}},xtick={1,2},
            xtick align=inside,
            xticklabel style={font=\small},
            ylabel={Area ($\mu m^2$)},
            ylabel near ticks,
            bar width = 7pt,
            xmin=0.6,
            xmax=2.5,
            ymin=1650,
            ymax=1800,
            legend style={at={(0.55,1.26)},
            draw=none,anchor=north,legend columns=-1},
        ]
        \addplot +[ybar, fill=mygrey,   draw=black, area legend] table [x={alpha},  y={baseline}] {diff-area.dat};
        \addplot +[      fill=mygreen, draw=black, area legend] table [x={alpha},  y={UCB}] {diff-area.dat};
        \addplot +[      fill=myyellow, draw=black, area legend] table [x={alpha},  y={EI}] {diff-area.dat};
        \addplot +[      fill=mymiddle, draw=black, area legend] table [x={alpha},  y={POI}] {diff-area.dat};
        \legend{Baseline,UCB,EI,POI}
    \end{axis}
\end{tikzpicture}} \hspace{.1in}
    \subfloat[]{\pgfplotsset{
    width=7.8cm,
    height=4.2cm
}

\begin{tikzpicture}
    \begin{axis}[
            ybar,
            xticklabels={\texttt{DesignWare}, \texttt{Kogge-Stone}},xtick={1,2},
            xtick align=inside,
            xticklabel style={font=\small},
            ylabel={Delay ($ps$)},
            ylabel near ticks,
            bar width = 7pt,
            xmin=0.6,
            xmax=2.5,
            ymin=300,
            ymax=355,
            legend style={at={(0.55,1.26)},
            draw=none,anchor=north,legend columns=-1},
        ]
        \addplot +[ybar, fill=mygrey,   draw=black, area legend] table [x={alpha},  y={baseline}] {diff-delay.dat};
        \addplot +[      fill=mygreen, draw=black, area legend] table [x={alpha},  y={UCB}] {diff-delay.dat};
        \addplot +[      fill=myyellow, draw=black, area legend] table [x={alpha},  y={EI}] {diff-delay.dat};
        \addplot +[      fill=mymiddle, draw=black, area legend] table [x={alpha},  y={POI}] {diff-delay.dat};
    \end{axis}
\end{tikzpicture}}
    \caption{Results with different acquisition functions.
    }
    \label{fig:diff-acq}
\end{figure}
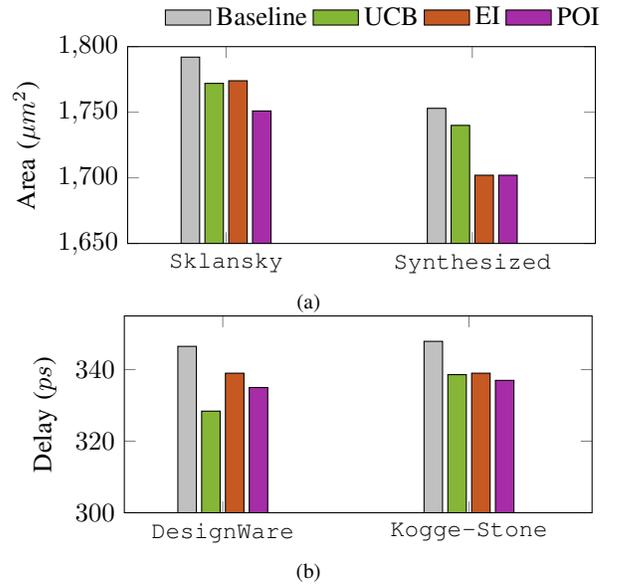

\begin{table*}[!tb]
	\centering
	\caption{Performance comparison between BO and baseline settings on multiple objectives}
	
	\label{tab:bl-bo-multi}
	\resizebox{17cm}{!}{
    \renewcommand{\arraystretch}{1.2}{
	\begin{tabular}{c|ccc|ccc}
		\toprule
		Adder structure     &\multicolumn{3}{c|}{Baseline}&\multicolumn{3}{c}{BO} \\ 
		\cline{2-7}       & Area ($\mu m^2$) & Energy ($fJ/op$) & Delay ($ps$) & Area ($\mu m^2$) & Energy ($fJ/op$) & Delay ($ps$) \\ 
		\hline
		\texttt{DesignWare}  & 2531 & 8160 & 346 & 2473 & 8170 & 345 \\
		\texttt{Sklansky}    & 1792 & 6100 & 356 & 1791 & 5920 & 350 \\
		\texttt{Kogge-Stone} & 2563 & 8780 & 347 & 2531 & 7980 & 340 \\
		\texttt{Synthesized} & 1753 & 5900 & 353 & 1754 & 5940 & 350 \\
		\hline
		Average           & 2160.1 & 7235.0 & 350.9 & 2137.9 & 7002.5 & 346.4 \\
		Ratio             & 1.0    & 1.0    & 1.0   & 0.989  & 0.967  & 0.987  \\
		\bottomrule
	\end{tabular}
    }
    }
\end{table*}

A set of complete scripts for adder synthesis is obtained from industry, which can provide us a reasonable good performance as baseline. 
We develop a Bayesian optimization engine using the proposed strategy, with synthesis tools integrated in the flow. 
It starts with a random initialization, and the maximum iteration number is set to be 30. 
Firstly, we target at a single objective such as area or energy or delay. 
The results are presented in \Cref{tab:bl-bo-single}. 
It can be seen that the Bayesian optimization can surpass the baseline performance within the limited budget.

Then we investigate the impacts of different acquisition functions. 
The results are demonstrated in \Cref{fig:diff-acq}. 
It can be observed that all acquisition functions can improve the performance compared to baseline settings, among which POI can lead to better performance in general.

Next we perform Bayesian optimization for multi-objective optimization. 
Since the acquisition function is defined over single value, we need to scalarize the output vector. 
In this experiment, the multiple objectives are merged into a single value by taking the weighted sum.
The results are demonstrated in \Cref{tab:bl-bo-multi}, from which we can observe that the Bayesian optimization can achieve comparable or better results.

\subsection{Results Comparison with Meta-heuristic Search}

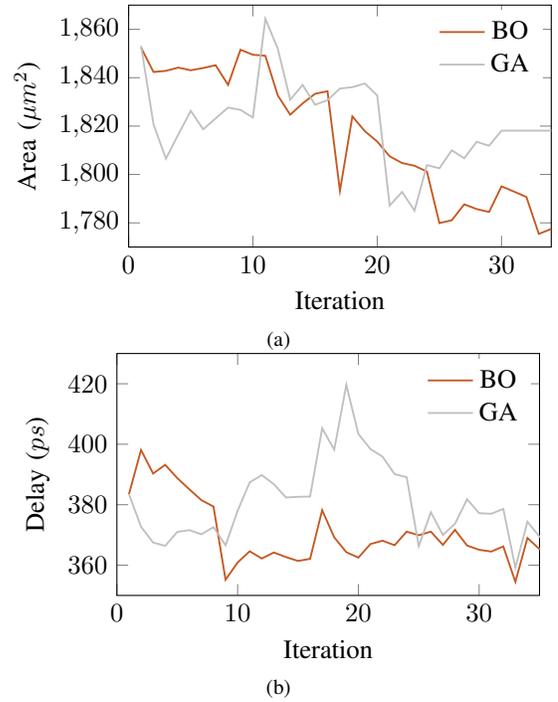
\begin{figure}[!tb]
	\centering
	
	\subfloat[]{\pgfplotsset{
    width=7.2cm,
    height=4.8cm
}
\begin{tikzpicture}[scale=1]
\begin{axis}[minor tick num=0,
xmin=0, xmax=34,
ymin=1770, ymax=1870,
xtick distance=10,
yticklabel style={/pgf/number format/.cd, fixed, fixed zerofill, precision=0, /tikz/.cd},
xlabel={Iteration},
ylabel={Area ($\mu m^2$)},
xlabel near ticks,
legend style={
  draw=none,
  at={(0.98,0.98)},
  anchor=north east,
  legend columns=1,
}
]
\addplot +[line width=0.7pt] [color=myyellow,   no marks]  table [x={alpha},  y={sklansky}] {converge-area.dat};
\addplot +[line width=0.7pt] [color=mygrey,   no marks] table [x={alpha},  y={sklansky-ga}] {converge-area.dat};
\legend{BO, GA}
\end{axis}
\end{tikzpicture}} \hspace{.2in}
	\subfloat[]{\pgfplotsset{
    width=7.2cm,
    height=4.8cm
}
\begin{tikzpicture}[scale=1]
\begin{axis}[minor tick num=0,
xmin=0, xmax=35,
ymin=350, ymax=430,
xtick distance=10,
yticklabel style={/pgf/number format/.cd, fixed, fixed zerofill, precision=0, /tikz/.cd},
xlabel={Iteration},
ylabel={Delay ($ps$)},
xlabel near ticks,
legend style={
  draw=none,
  at={(0.98,0.98)},
  anchor=north east,
  legend columns=1,
}
]
\addplot +[line width=0.7pt] [color=myyellow,   no marks]  table [x={alpha},  y={BO}] {converge-delay-irrg.dat};
\addplot +[line width=0.7pt] [color=mygrey,   no marks] table [x={alpha},  y={GA}] {converge-delay-irrg.dat};
\legend{BO, GA}
\end{axis}
\end{tikzpicture}}
	\caption{Evolving curve of the adder performance.}
    \label{fig:converge}
\end{figure}

Generally, the CAD tool design space exploration problem can be treated as a black-box function optimization. 
For black-box function optimization, there are a few well-known and commonly leveraged evolutionary algorithms, e.g., genetic algorithm or simulated annealing.
Thus, it is necessary to conduct comparison between the Bayesian optimization and the genetic algorithm. 
We plot the convergence curves for both Bayesian optimization and genetic algorithm on the same criterion during the optimization process, as shown in \Cref{fig:converge}. 
It can be seen that the Bayesian optimization can reach lower values for both criteria. 
Besides, the search trajectory of the genetic algorithm is not as stable as that of Bayesian optimization. 
Therefore, the Bayesian optimization shows significant superiority to the genetic algorithm.

\subsection{Discussion}

\begin{figure*}[!tb]
    \centering
    \subfloat[]{\pgfplotsset{
    width =7.5cm,
    height=4.3cm
}

\begin{tikzpicture}
\begin{axis}[
ybar,
xticklabels={Area, Energy, Delay},xtick={1,2,3},
xtick align=inside,
xticklabel style={font=\small},
ylabel={Ratio},
ylabel near ticks,
bar width = 7pt,
xmin=0.6,
xmax=3.5,
ymin=0.95,
ymax=1.04,
legend style={at={(1.1,1.26)},
	draw=none,anchor=north,legend columns=-1},
]
\addplot +[ybar, fill=mygrey,   draw=black, area legend] table [x={alpha},  y={Scale-up}] {multi-obj-scale-designware.dat};
\addplot +[      fill=mygreen2, draw=black, area legend] table [x={alpha},  y={Scale-down}] {multi-obj-scale-designware.dat};
\legend{Scale-up , Scale-down}
\end{axis}
\end{tikzpicture}}
    \hspace{-1.1in}
    \subfloat[]{\pgfplotsset{
    width =7.5cm,
    height=4.3cm
}
\begin{tikzpicture}
\begin{axis}[
ybar,
xticklabels={Area, Energy, Delay},xtick={1,2,3},
xtick align=inside,
xticklabel style={font=\small},
ylabel={Ratio},
ylabel near ticks,
bar width = 7pt,
xmin=0.6,
xmax=3.5,
ymin=0.93,
ymax=1.019,
legend style={at={(0.5,1.2)},
	draw=none,anchor=north,legend columns=-1},
]
\addplot +[ybar, fill=mygrey,   draw=black, area legend] table [x={alpha},  y={Scale-up}] {multi-obj-scale-ks.dat};
\addplot +[      fill=mygreen2, draw=black, area legend] table [x={alpha},  y={Scale-down}] {multi-obj-scale-ks.dat};
\end{axis}
\end{tikzpicture}}
    
    \subfloat[]{\pgfplotsset{
    width =7.5cm,
    height=4.3cm
}

\begin{tikzpicture}
\begin{axis}[
ybar,
xticklabels={Area, Energy, Delay},xtick={1,2,3},
xtick align=inside,
xticklabel style={font=\small},
ylabel={Ratio},
ylabel near ticks,
bar width = 7pt,
xmin=0.6,
xmax=3.5,
ymin=0.93,
ymax=1.015,
legend style={at={(0.5,1.2)},
	draw=none,anchor=north,legend columns=-1},
]
\addplot +[ybar, fill=mygrey,   draw=black, area legend] table [x={alpha},  y={Scale-up}] {multi-obj-scale-sklansky.dat};
\addplot +[      fill=mygreen2, draw=black, area legend] table [x={alpha},  y={Scale-down}] {multi-obj-scale-sklansky.dat};
\end{axis}
\end{tikzpicture}}
    \subfloat[]{\pgfplotsset{
    width =7.5cm,
    height=4.3cm
}

\begin{tikzpicture}
\begin{axis}[
ybar,
xticklabels={Area, Energy, Delay},xtick={1,2,3},
xtick align=inside,
xticklabel style={font=\small},
ylabel={Ratio},
ylabel near ticks,
bar width = 7pt,
xmin=0.6,
xmax=3.5,
ymin=0.93,
ymax=1.018,
legend style={at={(0.5,1.2)},
	draw=none,anchor=north,legend columns=-1},
]
\addplot +[ybar, fill=mygrey,   draw=black, area legend] table [x={alpha},  y={Scale-up}] {multi-obj-scale-irrg.dat};
\addplot +[      fill=mygreen2, draw=black, area legend] table [x={alpha},  y={Scale-down}] {multi-obj-scale-irrg.dat};
\end{axis}
\end{tikzpicture}}
    
    \caption{
        The performance comparison of various designs with different scaling methods.
        (a) \texttt{DesignWare}; (b) \texttt{Kogge-Stone}; (c) \texttt{Sklansky}; (d) \texttt{Synthesized}.
    }
    \label{fig:scaling}
\end{figure*}
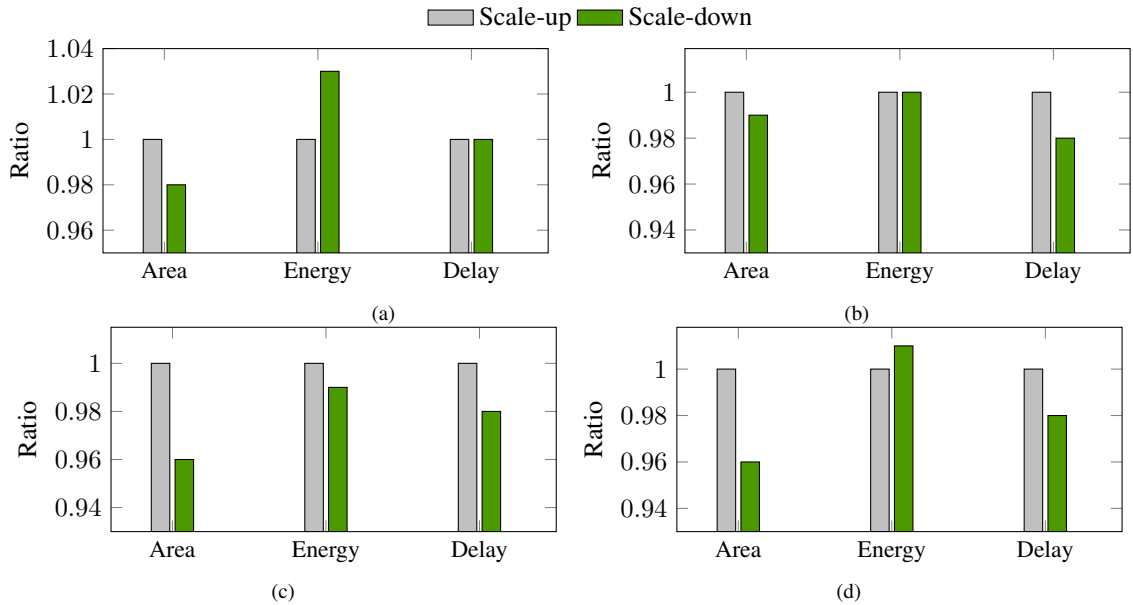

There are a few tips being observed during the experimental process, which have different effects on the final performance. 
Regarding the multi-objective optimization, we need to merge multiple objectives into a single scalar value. 
It should be noted that the values we obtained are not in the same scale, depending on the unit defined for each criterion. 
In order to avoid one objective value is dominated by another, scaling is considered necessary for those ``small" values to ensure that all the values contribute closely to the weighed objective. 
During the experiments it is found that scaling up smaller values and scaling down large values may lead to different results. 
\Cref{fig:scaling} shows the performance achieved using different scaling strategies. 
From the convergence points, most of the time scaling down larger values can achieve better results.

\section{Conclusion}
\label{sec:conclu}
Bayesian optimization is a machine learning approach which can be applied for better design. 
In this practice, we adapt Bayesian optimization for multi-objective optimization to simultaneously minimize the PPA values of a design. 
In our experiments, BO substantially outperforms typical evolutionary algorithms. 
According to our study and experimental practice, there are still plenty of room for improvement. 
First, the design spaces on the front-end and back-end are separated, which may still lead to local optimal points. 
A natural way for improvement is to extend the design space to a unified exploration framework. 
The second thing is that we deal with multi-objective by simply using scalarization, which requires tuning efforts and tricks. 
A more elegant way to handle this would be of great help. 

\section*{Acknowledgments}
This work is supported by The Research Grants Council of Hong Kong SAR (Project No.~CUHK24209017).

\balance
\bibliographystyle{IEEEtran}
\bibliography{../ref/Top-sim,../ref/DSE,../ref/Software,../ref/LEARN,../ref/CAD,../ref/DFM,../ref/TEST,../ref/FPGA}

\end{document}